\documentstyle[preprint,aps]{revtex}

\begin{document}

\title{
Influence of reptations on conformations of the homopolymer
in Monte Carlo simulation
}

\author{
Yuri A. Kuznetsov, Edward G. Timoshenko\thanks{Corresponding author. 
E-mail: Edward.Timoshenko@ucd.ie; Web page: http://darkstar.ucd.ie;
Phone: +353-1-7062821; Fax: +353-1-7062127
 }
}

\address{
Theory and Computation Group, Department of Chemistry,\\
University College Dublin, Belfield, Dublin 4, Ireland
}

\maketitle

\begin{abstract}
We study the role of topological restrictions for the conformational
structure of a nonphantom homopolymer chain in a lattice Monte Carlo
model. In the athermal regime we find that the standard Metropolis
algorithm violates the detailed balance condition if both local
monomer and global reptational moves are included. However, if about
$1$ reptation is performed per $N$ local moves the balance is recovered
and we obtain the Flory exponent value close to $\nu=0.588$, where $N$
is the degree of polymerisation.
We also find that the structure of the collapsed globule is
different at equilibrium from that after the late stage of folding kinetics.
Namely, due to reptations the end, and even more so, the penultimate monomer
groups tend to be buried inside the globule core with other monomers thus
being more exposed to the surface.
\end{abstract}


\section{Introduction}
\label{sec:intro}

It was early recognised that topological restrictions play an important
role in determining the conformation and dynamics of polymer solutions
\cite{Books}. One of the most important topological restrictions is
the integrity of chain links,
which implies that different parts of the polymer chain cannot pass
through each other.  Unfortunately, it is rather difficult to include
such restrictions to analytical theories, especially to simple
mean--field ones.

Nevertheless, the effects of topological restrictions can be properly
studied by computer simulation techniques \cite{Binder-book}.
One such technique developed by the authors in Ref. \cite{CoplmMonte}
is based on a lattice Monte Carlo model.
Here the Monte Carlo updates scheme includes local monomer and reptational
chain moves. The former is an attempt to move a randomly
chosen monomer to a randomly chosen nearby lattice site.
The latter is an attempt to move a randomly chosen end monomer
to a randomly chosen lattice side near the other polymer end.
It is important to emphasise that this model does not permit
moves that would violate the integrity of links,
i.e. the chain is strictly nonphantom.
This is ensured automatically due to the particular choice of the
connectivity and the excluded volume parameters.

In this paper we study how the probability of performing reptational
moves affects the conformations of a single homopolymer chain at
good and poor solvent conditions.
In particular, we compare the globules obtained by
a kinetic process and at true equilibrium.
Since the kinetics of polymer collapse proceeds
through formation and growth of locally collapsed clusters along
the chain, with their final unification into a single globule, one
may expect that after the shape optimisation stage  
the homopolymer globule possesses a comparatively simple 
topological structure \cite{CoplmMonte}. Indeed, a typical
coil conformation before the quench possesses a statistically
small number of entanglements, or knots, and the folding
kinetics adds virtually nothing to that number.
Further relaxation of the globule towards the equilibrium requires  
participation of the chain ends and, thus, can be viewed as 
an auto--reptational stage.
This auto--reptation time can be estimated as,
$\tau_{rept} = \tau_0  N^3$ \cite{GrosbKhokh-book},
which can yield a time of order $10^3$ sec under usual
experimental conditions \cite{GrosbKinet}.

\section{The Good Solvent Regime}
\label{sec:flory}

In this section we shall test the Monte Carlo scheme 
by determining the value of the swelling exponent $\nu$ \cite{Books},
in the good solvent regime.

Generally, it seems that to improve the convergence of the system to
equilibrium various types of global moves, such as reptations,
may be included in addition to local ones.
The resulting equilibrium state should not depend on a 
particular scheme involved. To test this assumption
let us study how the probability of performing reptations, $p_r$,
affects the value of the Flory exponent of the coil.
The results are presented in Tab. \ref{tab:floryRg2}.
One can see that the value of $\nu$ obtained in the 
scheme without reptations (at the bottom of the 
left column) is somewhat higher than even the mean--field 
prediction $\nu=3/5$. As we increase the probability of performing reptations
$\nu$ starts to decrease. In the limit when only reptations are performed, 
$p_r=1$, the swelling exponent value
is found to be $0.56$.
Note that for the probability $p_r = 1/N$ the measured swelling exponent
is remarkably close to the most accurate result obtained from the
renormalisation group theory, $\nu \approx 0.588$ \cite{Books}.

We may conclude therefore that the scheme with only reptations involved
leads to more compact entangled conformations, resulting in
underestimation of the swelling exponent.
On the other hand, the scheme with local monomer moves only favours 
topologically simple conformations as it is rather improbable to create
a knot by local movements.
If some knots already exist in the initial conformation,
the local monomer movements would tend to disentangle them.
Indeed, one can imagine that simple `shaking' of an entangled
boot strap would more likely disentangle it rather than entangle it more.
This weak topological effect reduces the number of entanglements,
which leads to a larger radius of gyration and overestimation of
the swelling exponent in the scheme with local moves only.

Such strong dependence on the reptation probability $p_r$ is quite
unexpected from the point of view of the standard Monte Carlo paradigm.
Of course, the Metropolis check in itself is not sufficient
for satisfying the detailed balance condition.
One also has to ensure the condition that the phase space of the
system is sampled uniformly by attempted Monte Carlo moves \cite{Parisi-book}
(see Appendix for more detail).
Although, this may be quite simple to ensure for point--like objects,
in our case of a nonphantom chain this is not so.
The above observations do indicate clearly that the current
sampling procedure is not uniform, but biased.
The bias is present in both schemes with reptations only and
local moves only, but has the opposite effect.
We have also seen that if about $1$ reptation
is performed per $N$ Monte Carlo steps
the topological effects of entanglements and disentanglements
balance each other, making the sampling of the phase space
essentially uniform and producing the correct swelling exponent.

However, we should emphasise that this problem of the improper
influence of reptations is only present for the Flory
coil and it is irrelevant for the ideal coil. 
Both schemes with local moves only and reptations
only would give $\nu = 1/2$ for the ideal solution.
That is why reptation techniques are extremely popular
and well justified for studying melts and concentrated
solutions, in which reptations also may be the only
physically relevant motions.
We should also emphasise that if the chain was phantom
we would not have this problem either.

Even though the above discussed effect for the Flory coil
is clearly of topological origin, it only presents a problem for
the simulation procedure and has no implications for real polymers.

\section{Structure of the Homopolymer Globule}
\label{sec:homoglob}

In this section we shall study the structural differences between the two
globules of an open homopolymer: one with a relatively simple topological
structure corresponding to the late stage of folding kinetics, and
another with topological entanglements. In practice these
simulations have been carried out in the following way.
Using the lattice Monte Carlo method a large set of
homopolymer globules was produced by independent kinetic processes
starting from initial coil conformations.
Reptational moves were not included during this simulation.
To produce true equilibrium distribution for the
globule an additional simulation was applied to the
set with amount of reptational moves equal to $p_r = 1/N$.

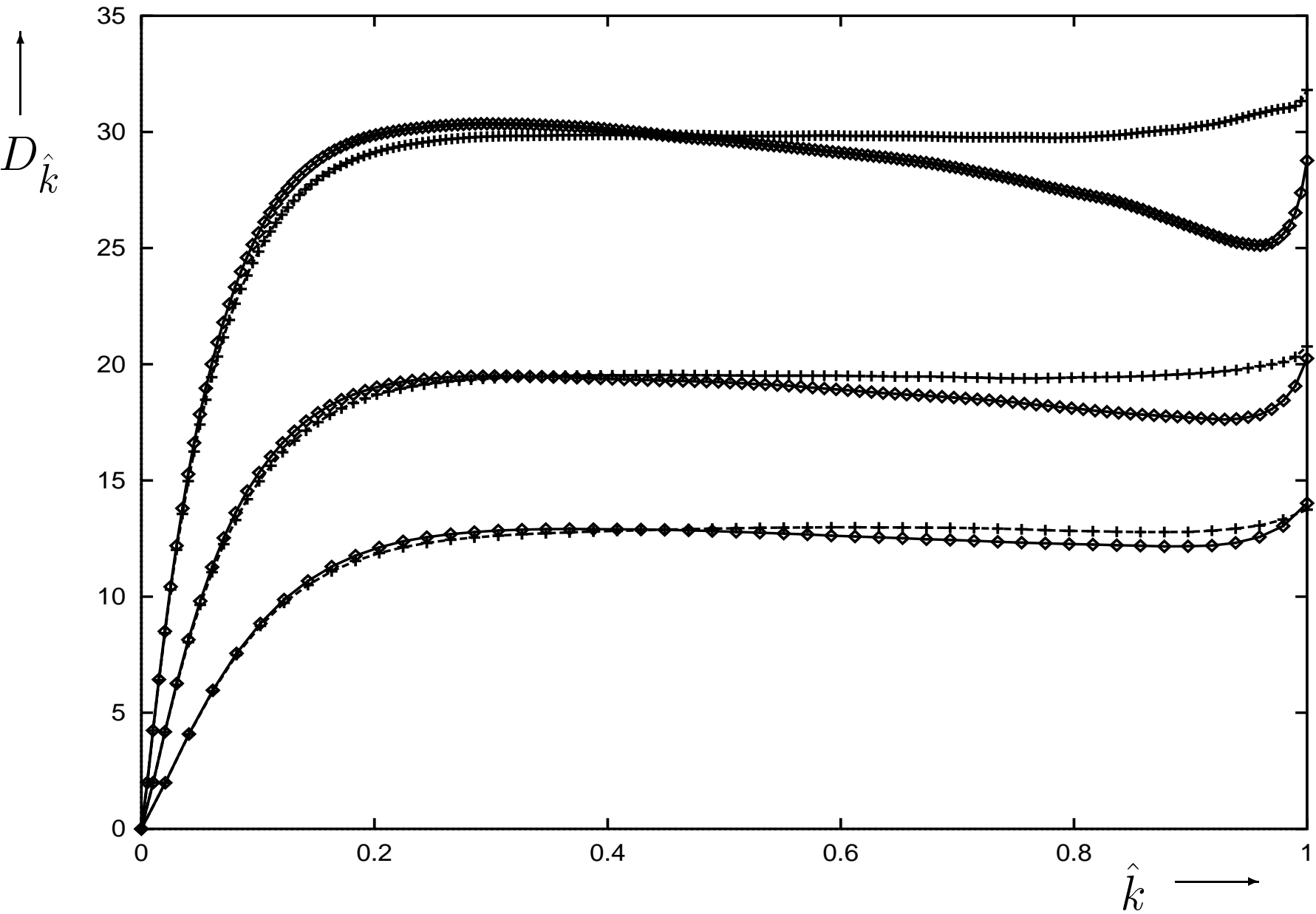
\begin{figure}
\caption{ \label{fig:glb_dk}
Plot of the partially summated mean squared distances, $D_{\hat{k}}$, vs
the normalised chain index, $\hat{k} = k/(N-1)$, for
homopolymer globules of different sizes and different
simulation procedures.
Pairs of lines correspond respectively to the following values of the
degree of polymerisation (from bottom to top): $N = 50$, $N = 100$
and $N = 200$. Lines denoted by diamonds and pluses in each pair
correspond to the simulation procedures with, $p_r = 1/N$,
and without reptations, $p_r = 0$, respectively.
}
\end{figure}

Let us introduce the mean squared distances along the chain, $D_{mn}$, and
their partially summated combinations, $D_k$
\begin{equation}
D_{mn} = \left\langle ({\bf X}_m - {\bf X}_n )^2 \right\rangle,
\qquad D_k \equiv \frac{1}{N - k} \sum_{i = 0}^{N-1-k} D_{i\,i+k}.
\label{glob:dk}
\end{equation}
In Fig.~\ref{fig:glb_dk} we present the quantity
$D_{\hat{k}}$ versus the normalised chain index, $\hat{k}$,
for polymers of different degrees of polymerisation.
For very small chain indices function $D_k$ does not depend on
$N$ and on the particular simulation procedure, which reflects
local packing of monomers in the dense globule.
Let us consider first the behaviour of this function
for globules prepared without reptations.
For small values of the chain index the function is almost linear,
up to some cross--over value that scales as $N^{2/3}$.
Then it saturates to some level, which is proportional to
the radius of the globule, thus also scaling as $N^{2/3}$.
Interestingly enough, for values of the chain index in the vicinity
of the chain ends the function $D_k$ increases once again.
This phenomenon actually reflects the mechanism of
the polymer collapse during late coarsening stage.
The globule is usually formed by a final unification of two
end clusters and, sometimes, a few middle clusters 
(see e.~g. Figs.~4, 6 and 10 in Ref.~\onlinecite{CoplmMonte}).
Thus, the chain ends possess a somewhat higher probability
to appear on opposite sides of the globule than the rest
of monomers. This effect in the mean squared distances
is fairly weak as the function experiences about $10\%$
increase towards the ends.

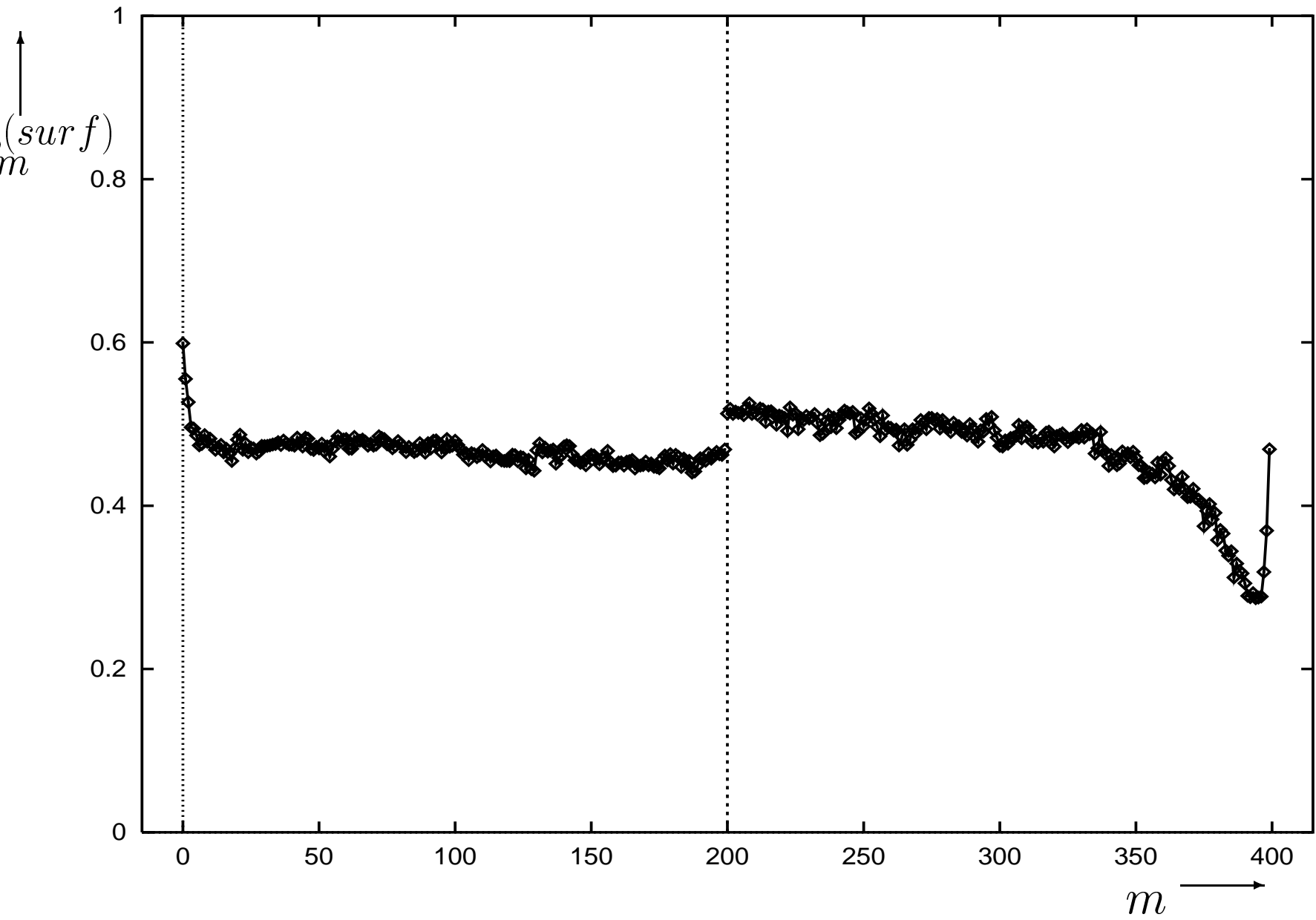
\begin{figure}
\caption{ \label{fig:glb_surf}
Plot of the probability, $P^{(surf)}_m$, for the $m$-th monomer
in the chain to appear on the globule surface vs the monomer index,
$m$, for the homopolymer with the degree of polymerisation, $N = 400$,
for different simulation procedures.
The left-- and right--hand--side curves correspond to the
kinetic (no reptations) and reptational simulation procedures.
Both distributions are symmetric in monomer index, $m$, and for convenience
they are presented only on halves of the interval.
}
\end{figure}

In fact, this observation is related to the observation that the chain ends
are more exposed to the globule surface, and it can be better
justified by considering the probability of the $m$-th
monomer in the chain to appear on the surface of the
globule, $P^{(surf)}_m$. This is presented in the
left--hand--side part of Fig.~\ref{fig:glb_surf}.
Thus, the function $P^{(surf)}_m$ for kinetic simulation is
nearly constant except for a few monomers at the very chain ends.
In particular, for the end monomer this probability is about
$1.3$ times larger than that for a monomer in the centre of the chain.

The effect due to applying reptational moves during later kinetic
stages is quite distinguishable in observables in
Figs.~\ref{fig:glb_dk} and \ref{fig:glb_surf}.
For sufficiently large values of the chain indices,
$\hat{k} \gtrsim 0.35-0.4$, the function of mean squared
distances obtained from simulation with reptations
(see lines denoted by diamonds in Fig.~\ref{fig:glb_dk})
lies below the appropriate curves from the kinetic simulation.
The situation is just the reverse for smaller $k$, with the
difference slowly vanishing towards $k = 0$.
Thus, the reptational curve for $D_{k}$ behaves in the following way:
it is linear for small $k$, reaches a maximum at around $\hat{k} = 0.3$
and then slowly decreases reaching a minimum in the vicinity of the chain ends.
At the very end the function increases strongly, still being
significantly smaller that for the function $D_k$ without reptations.

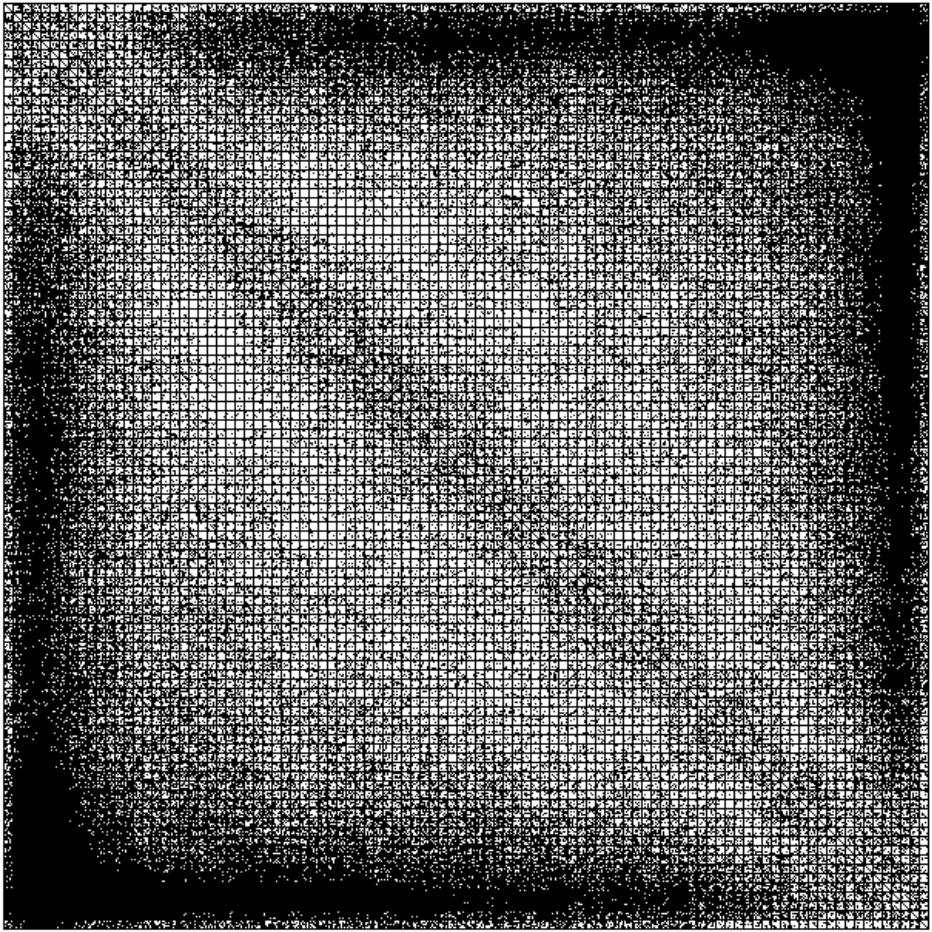
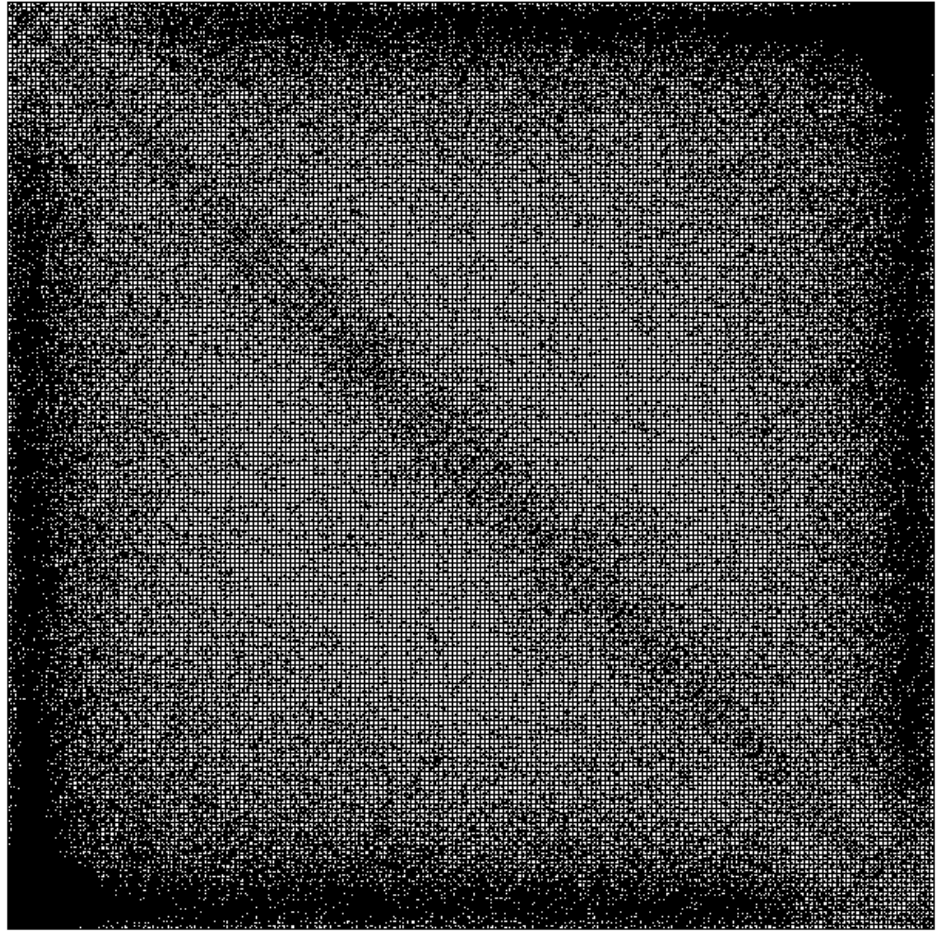
\begin{figure}
\caption{ \label{fig:glb_dmn}
Diagrams of the ratios of the mean squared distances with and
without reptations, $D_{mn}^{(r)} / D_{mn}^{(n)}$ for
different values of the degree of polymerisation.
Diagrams (a) and (b) correspond respectively to the following values of the
degree of polymerisation: $N = 100$ and $N = 200$.
Indices $m$, $n$ start counting from the upper
left corner. Each matrix element is denoted by a quadratic
cell with varying degree of black colour, the darkest and the
lightest cells corresponding respectively to the smallest and to the largest
ratios of the mean squared distances. Intensity of black colour
in the near diagonal elements corresponds approximately to the ratio
$D_{mn}^{(r)} / D_{mn}^{(n)} = 1$.
}
\end{figure}

Note that the partially summated mean squared distances are not fully
informative due to the role of the end effects in an open chain.
Thus, let us consider the mean squared distances
as functions of two indices. In Fig.~\ref{fig:glb_dmn} we exhibit the
ratios of mean squared distances with and without reptations,
$D_{mn}^{(r)} / D_{mn}^{(n)}$. Depending on the behaviour
of this ratio, the monomers in the chain may be roughly divided
into three groups:
for monomers in the centre of the chain,
$0.1 \lesssim m/N \lesssim 0.9$, the mean squared
distances significantly increase due to reptations;
for groups of penultimate to the end monomers,
$3 \lesssim m,\, N-m-1 \lesssim 0.1 N$, the distances
decrease due to reptations, with the effect becoming more
pronounced on approaching the ends of the chain;
and for a few end monomers the distances decrease
but much more weakly. Thus, we can conclude that reptations push the
end groups somewhat and the penultimate groups of monomers more so
towards the centre of the globule, thus reducing their
mean squared distances between each other and monomers from
the central group. As the density of the
globule does not change here this leads to
an increase of the mean squared distances for monomers
in the centre of the chain.

Such behaviour of the mean squared distances is quite consistent
with the plot of the surface probability $P^{(surf)}_m$ after
reptational stage (right--hand--side of Fig.~\ref{fig:glb_surf}).
This quantity increases for monomers in the central group by about
$10 \%$ and, instead of being constant, decreases slowly towards
the ends of the chain. The drop in the probability is most pronounced
for monomers from the penultimate groups.
The function rapidly increases for a few monomers at
the very ends of the chain, although, they still possess
a lower probability to be found on the surface than
monomers from the central group.
This rapid increase of the probability for a few end monomers
may be interpreted as an effect of a single
end seeking to maximise its entropy.
Indeed, the chain ends are most free to explore the surface,
whilst the rest of monomers are more restricted due to connectivity.

\section{Conclusion}

In this paper we have applied the lattice Monte Carlo model
of Ref. \onlinecite{CoplmMonte} to study the role of reptations
for conformations of the nonphantom homopolymer chain.

First, we have considered the athermal good solvent regime
and discovered that the frequency of performing reptations
significantly affects the size and even the swelling exponent
of the polymer.
This problem arises due to a nonuniform sampling of the phase space
of the system by attempted Monte Carlo moves.
For a nonphantom chain it is not clear how to ensure
uniform sampling especially between conformations with different
topological numbers. The scheme with reptations only is biased towards
entangled conformations, while the scheme with local moves only samples
topologically simple states more.
However, the balance is recovered when the probability
of reptations is equal to $p_r = 1/N$.

Second, we have compared structures of the two globules: one corresponding
to late stages of kinetics and another to the true equilibrium.
The globule after the late kinetic stage is characterised by
a uniform monomer distribution except for the end monomers,
which have a higher probability to be on the surface.
Due to reptations the end, and even more so, the penultimate monomer
groups are pushed more towards the centre of the globule, with
the central monomer group thus being more exposed to the surface.
A typical magnitude of this effect in observables expressing
the $k$-dependent properties is about a few dozens percents in the relative
change, while the global characteristics such as the mean energy and
radius of gyration are practically unaffected by this conformational change.
The latter is a physically important conclusion for real polymers.

\acknowledgments

The authors are grateful for interesting discussions to
Professor K.A.~Dawson, Professor F.~Ganazzoli, and especially to
Professor A.Yu.~Grosberg.
We also acknowledge the support from Enterprise Ireland
grants SC/99/186 and IC/1999/01 and the support of the Centre for High
Performance Computer Applications, UCD.

\appendix

\section*{The problem of uniform sampling in the Metropolis algorithm}

The system is simulated based on 
the Metropolis algorithm \cite{Binder-book} for calculation of
the statistical averages in the system at temperature $T$.

Its main idea is to construct a Markov process
in the phase space,
\begin{equation}
{\bf X}_A (1) \rightarrow {\bf X}_A (2) \rightarrow \ldots \rightarrow
{\bf X}_A (i-1) \rightarrow {\bf X}_A (i) \rightarrow \ldots,
\label{lmc:MarChain}
\end{equation}
where each state ${\bf X}_A (i)$ follows from the previous one
${\bf X}_A (i-1)$ according to the transition probability,
$\Pi ( {\bf X}_A' \rightarrow {\bf X}_A )$.

Let $P_0(\bbox{X}_A)$ denotes the initial probability with any 
distribution, e.g. uniform.
At each time step we update these probabilities by a certain rule.
The updated probabilities set may be obtained as,
$P_1(\bbox{X}_A) = \sum_{\bbox{X}'_A} \Pi(\bbox{X}'_A\rightarrow 
\bbox{X}_A)\, P_0(\bbox{X}'_A)$,
where $\Pi(\bbox{X}'_A\rightarrow \bbox{X}_A)$ is the corresponding
transition probability.
We then repeat the above procedure many times using the results
from the previous step as input.

Let us introduce
the {\it balance condition}, 
which expresses the stationarity of the equilibrium
distribution, $P_{eq}({\bf X}_A)= Z^{-1} \exp(-\beta H({\bf X}_A)),$ with
respect to the action of the Markov process:
\begin{equation} \label{Bal}
P_{eq}({\bf X}_A) = \sum_{{\bf X}'_A} \Pi({\bf X}'_A\rightarrow {\bf X}_A)
\,P_{eq}({\bf X}'_{A'}).
\end{equation}
Next,
the {\it ergodicity condition} means that the phase space does
not factorise into disjoint parts, i.e.
for any ${\bf X}'_A$ and ${\bf X}_A$ there exists a finite sequence of
${\bf X}_A(i)$ $i=1,\ldots, m$ such that 
$\Pi({\bf X}'_A \rightarrow {\bf X}_A)
\not= 0$ and ${\bf X}_A(1)={\bf X}'_A$ an ${\bf X}_A(m+1)={\bf X}_A$.
In other words, the system can go from any state to any other state
by a finite number of steps.

These two conditions are sufficient
for convergence of the ensemble produced by a Markov process
to the equilibrium distribution \cite{Parisi-book}. 
To show that we need to
introduce the notion of the ``distance'' between two
ensembles $E$ and $E'$ via the norm \cite{Parisi-book},
\begin{equation}
|| E-E' || = \sum_{{\bf X}_A} |P({\bf X}_A) - P'({\bf X}_A)|. 
\end{equation}
Now, if the ensemble $E'$ is obtained from $E$ by action of one step in the
Markov chain we have,
\begin{equation}
P'({\bf X}_A) =\sum_{{\bf X}'_A} \Pi({\bf X}'_A\rightarrow {\bf X}_A)\, 
P({\bf X}'_A).
\end{equation}
This allows us to obtain the following estimate,
\begin{eqnarray}
|| E' - E_{eq} || &=& \sum_{{\bf X}_A} \left| \sum_{{\bf X}'_A}
\Pi({\bf X}'_A \rightarrow {\bf X}_A) (P({\bf X}'_A)-P_{eq}({\bf X}'_A))
 \right|  \label{estI}\\
& \leq & \sum_{{\bf X}_A,{\bf X}'_A} \Pi({\bf X}'_A\rightarrow {\bf X}_A)\,
|P({\bf X}'_A) - P_{eq}({\bf X}'_A)| = || E- E_{eq} ||, \nonumber
\end{eqnarray} 
where we have used the non--negativeness of the transition probability 
and the normalisation condition,
$\sum_{{\bf X}'_A} \Pi({\bf X}'_A \rightarrow {\bf X}_A)=1$.
Thus, application of such a step moves the ensemble closer to the
equilibrium.
Finally, note that the inequality in Eq. (\ref{estI}) becomes strict thanks
to the ergodicity, and so the equality is only possible when the ensemble
has reached the equilibrium.

Instead of the balance condition (\ref{Bal}) one can apply
a more restrictive {\it detailed balance condition},
which relates the transition probabilities
of the forward and backward transitions,
\begin{equation} \label{lmc:det_bal} 
\exp\left( -\frac{H [{\bf X}_A]}{k_B T} \right) \, 
\Pi ( {\bf X}_A \rightarrow {\bf X}_A' ) = 
\exp\left( -\frac{H [{\bf X}'_A]}{k_B T} \right)
\, \Pi ( {\bf X}_A' \rightarrow {\bf X}_A ),
\end{equation}
which obviously produces the latter balance
condition by summing over ${\bf X}'_A$.

In the Metropolis algorithm one step of the Markov chain is generated
in two steps:\\
1. Given ${\bf X}_A(n)$ one generates a trial conformation
${\bf X}_A(T)$ by a random algorithm with a symmetric
transition probability,
\begin{equation}
\Pi_{S}({\bf X}_A(n) \rightarrow {\bf X}_A(T))=
\Pi_{S}({\bf X}_A(T) \rightarrow {\bf X}_A(n)).
\end{equation}
This process looks like a simple random walk in the
phase space and does not depend on the particular
Hamiltonian $H$ of the system.
For instance, for the 
Ising model it would mean to pick up a randomly chosen spin.\\
2. Perform a transition with the conditional
probability $\Pi_{M}({\bf X}_A(n) \rightarrow {\bf X}_A(T))$,
so that the total transitional probability is,
\begin{equation}
\Pi({\bf X}_A(n) \rightarrow {\bf X}_A(T))=
\Pi_{S}({\bf X}_A(n) \rightarrow {\bf X}_A(T))\,
\Pi_{M}({\bf X}_A(n) \rightarrow {\bf X}_A(T)).
\end{equation}
A simple choice of this transition probability was proposed by Metropolis 
et al,
\begin{equation}
\Pi_M ( {\bf X}_A \rightarrow {\bf X}_A' ) = \left\{
\begin{array}{cl}
\exp\left( -\frac{\Delta H} {k_B T} \right), \quad & \mbox{for}\ 
\Delta H \equiv H [{\bf X}_A'] - H [{\bf X}_A] > 0, \\
1, & \mbox{for}\ \Delta H \leq 0. \\
\end{array}
\right.
\label{lmc:Wex}
\end{equation}
Due to the explicit definition of $\Pi_M$ and symmetricity
of the matrix $\Pi_S$ the total transition
probability $\Pi$ still satisfies the detailed balance condition.

Thus, despite many popular beliefs, the Metropolis check
itself is not sufficient for satisfying the detailed balance condition.
We also have to ensure the condition that the phase space
of the system is sampled uniformly
by attempted Monte Carlo moves, so that the transition
probability $\Pi_S(old\rightarrow new)$ is symmetric.
This may be simple enough to ensure for point--like objects
by simply picking them at random, however,
in our case of a nonphantom polymer chain this is not quite so easy
due to the topology.

%%%%%%%%%%%%%%%%%%%%%%%%%%%%%%%%%%%%%%%%%%%%%%%%%%%%%%%%%%%%%%%%%%%%%

%%%%%%%%%%%%%%%%%%%%%%%%%%%%%  TABLES %%%%%%%%%%%%%%%%%%%%%%%%%%%%%%%%%%%%%%%

\begin{table}
\caption{ \label{tab:floryRg2}
Values of the mean  squared radius of gyration, $R_g^2$, vs the degree of
polymerisation, $N$, for different simulation procedures.
Here $Q$ is the number of statistical measurements, $p_r$ is the probability
of making reptations in the scheme, with $1 - p_r$ being
the probability of making local monomer moves. The exponents $\nu_1$ and
$\nu_2$ were obtained by a least square fit of $\log R_g$
vs $\log N$ in the ranges $100-1000$ and $500-1000$ respectively.
}
\vskip 2mm
\begin{tabular}{||r||c|c|c|c||}
%\hline
$N$& $p_r=0$&$p_r=1/N$&$p_r=0.1$& $p_r=1$\\
$Q$  & 80,000 & 60,000 & 40,000 & 40,000 \\
\hline
20   &  13.14 &  12.55 &  12.29 &  11.50 \\
30   &  22.21 &  20.74 &  19.88 &  18.44 \\
50   &  42.16 &  38.50 &  35.73 &  32.63 \\
70   &  64.89 &  57.66 &  52.54 &  48.45 \\
100  &  98.17 &  88.15 &  77.75 &  72.54 \\
150  &  162.2 &  142.7 &  122.6 &  112.9 \\
200  &  227.2 &  201.2 &  169.0 &  156.8 \\
300  &  381.2 &  322.7 &  261.8 &  245.9 \\
500  &  696.5 &  588.1 &  460.0 &  435.9 \\
700  &  1029  &  856.1 &  667.2 &  637.6 \\
1000 &  1643  &  1329  &  991.5 &  947.7 \\
\hline
$\nu_1$& 0.608 & 0.587 &  0.551 &  0.559 \\
$\nu_2$& 0.619 & 0.588 &  0.554 &  0.560 \\
%\hline
\end{tabular}
\end{table}

%%%%%%%%%%%%%%%%%%%%%%%%%%%%% FIGURES %%%%%%%%%%%%%%%%%%%%%%%%%%%%%%%%%%%%%%%

%%%%%%%%%%%%%%%%%%%%%%%%%%%%%%%%%%%%%%%%%%%%%%%%%%%%%%%%%%%%%%%%%

\end{document}